\documentclass[sigplan,screen]{acmart}
\usepackage{hyperref}

\AtBeginDocument{%
  \providecommand\BibTeX{{%
    \normalfont B\kern-0.5em{\scshape i\kern-0.25em b}\kern-0.8em\TeX}}}

\usepackage{multirow}
\usepackage{array}
\begin{document}
\newcommand\MyBox[2]{
  \fbox{\lower0.75cm
    \vbox to 1.7cm{\vfil
      \hbox to 1.7cm{\hfil\parbox{1.4cm}{#1\\#2}\hfil}
      \vfil}%
  }%
}

\def\colorModel{hsb} 

\newcommand\ColCell[1]{
  \pgfmathparse{#1<50?1:0}  
    \ifnum\pgfmathresult=0\relax\color{white}\fi
  \pgfmathsetmacro\compA{0}      
  \pgfmathsetmacro\compB{#1/100} 
  \pgfmathsetmacro\compC{1}      
  \edef\x{\noexpand\centering\noexpand\cellcolor[\colorModel]{\compA,\compB,\compC}}\x #1
  } 
\newcolumntype{E}{>{\collectcell\ColCell}m{0.4cm}<{\endcollectcell}}  
\newcommand*\rot{\rotatebox{90}}

\settopmatter{printacmref=false}
\setcopyright{none}
\renewcommand\footnotetextcopyrightpermission[1]{}
\pagestyle{plain}

\pagestyle{fancy}
\lhead{Cross-Domain Shopping and Stock Trend Analysis}
\chead{}
\rhead{Aditya Pandey, Haseeba Fathiya, Nivedita Patel}
\cfoot{\thepage}

\title{Cross-Domain Shopping and Stock Trend Analysis}


\author{Aditya Pandey}
\affiliation{%
  \institution{New York University}
  \city{New York}
  \country{USA}
}
\email{ap6624@nyu.edu}

\author{Haseeba Fathiya}
\affiliation{%
  \institution{New York University}
  \city{New York}
  \country{USA}
}
\email{hf2313@nyu.edu}

\author{Nivedita Patel}
\affiliation{%
  \institution{New York University}
  \city{New York}
  \country{USA}
}
\email{np2457@nyu.edu}


\begin{abstract}
This paper presents a cross-domain trend analysis that aims to identify and analyze the relationships between stock prices, stock news on Twitter, and users' behaviors on e-commerce websites. The analysis is based on three datasets: a US stock dataset, a stock tweets dataset, and an e-commerce behavior dataset. 
The analysis is performed using Hadoop\cite{hdfs}, Hive\cite{hive} and Tableau\cite{tableau}, which allows for efficient and scalable processing and visualizing of large datasets.

The analysis includes trend analysis of Twitter sentiment (positive and negative tweets) and correlation analysis, including the correlation between tweet sentiment and stocks, the correlation between stock trends and shopping behavior, and the understanding of data based on different slices of time. By comparing differe features from the datasets over time, we hope to gain insight into the factors that drive user behavior as well as the market in different categories.
The results of this analysis can provide valuable insights for businesses and investors to inform decision-making. 

We believe that our analysis can serve as a valuable starting point for further research and investigation into these topics.
\end{abstract}


\settopmatter{printacmref=false}


\maketitle

\section{Introduction}

In recent years, the intersection of the stock market and social media activity has become increasingly relevant for businesses and investors seeking to make informed decisions. The rise of social media and mobile phone usage has made it easier for people to share information and opinions about stocks, companies and the market in general, and this information can be valuable for businesses and investors looking to understand market trends and investor sentiment. In addition, the rapid growth of e-commerce and the varied behaviors of online shoppers have made it important for companies to understand their customers' needs and preferences in order to improve their products and services and develop effective marketing and advertising strategies. This information can also be useful for demand forecasting for businesses and manufacturers.

Sentiment analysis is a technique that can be used to extract and analyze the emotional or evaluative content of text data, such as social media posts or online reviews. In the context of stock market data, sentiment analysis can be used to analyze the sentiment of tweets about specific stocks or the market as a whole, in order to gain insights into market trends and investor sentiment. This type of analysis can be particularly useful for businesses and investors looking to make informed decisions about the stock market, as it provides a way to quantify and understand the sentiment of social media conversations about stocks. In this paper, we will examine the use of sentiment analysis to analyze twitter stock data and explore its potential for informing decision-making.

Despite the limited time-range over which we have analyzed the datasets, we believe that our analysis can provide useful insights for businesses and investors seeking to understand these relationships.

\section{Motivation}
As a business owner or data scientist, staying on top of market trends and consumer behavior is essential for success. That's where cross-domain shopping and stock trend analysis comes in. Analyzing these allows you to gain a better understanding of the correlations between shopping trends and stock trends, giving you valuable insights into consumer demand and market trends.

Beneficiaries of cross-domain shopping and stock trend analysis include businesses, advertising companies, and manufacturers. By gaining a deeper understanding of the relationships between different types of data, these stakeholders can better serve their customers and optimize their operations to drive growth and success.

Understanding customer behavior and needs are critical for companies, as it can help them evaluate customer satisfaction and make necessary improvements to their products and services. A centralized sentiment trend analysis system can provide more accurate and useful insights into customers'  experiences, thoughts, and beliefs. This can be beneficial for businesses and advertising companies, as it can help them improve their offerings and increase profits. 

In addition, through the analysis of stock trends and tweet sentiment trends, businesses will be able to react to any positive or negative trends in the market by offering new products and services that meet the needs and preferences of their customers. This can help businesses stay ahead of the curve and stay competitive in an increasingly fast-paced and dynamic market.

\section{Goodness}

In assessing the 'goodness' of the analysis, we considered several factors which led us to believe that the results of our analytics are correct and can be trusted 

First, the results of the analysis were compared to the general expectation of user behaviors during different periods of time. This allowed us to determine whether the results were in line with what we would expect to see based on our understanding of user behavior. 

Second, we checked to see whether the market-related results were consistent with the general understanding of basic stock behavior. For example, the fact that the market is closed on the weekends, was a factor in the result. 

Finally, we explored any interesting results in order to understand the reasoning behind them and determine whether they were significant or potentially due to some other factor(s).

\section{Datasets}
\label{sec:dataset}
\begin{figure*}[h]
  \includegraphics[width=\textwidth]{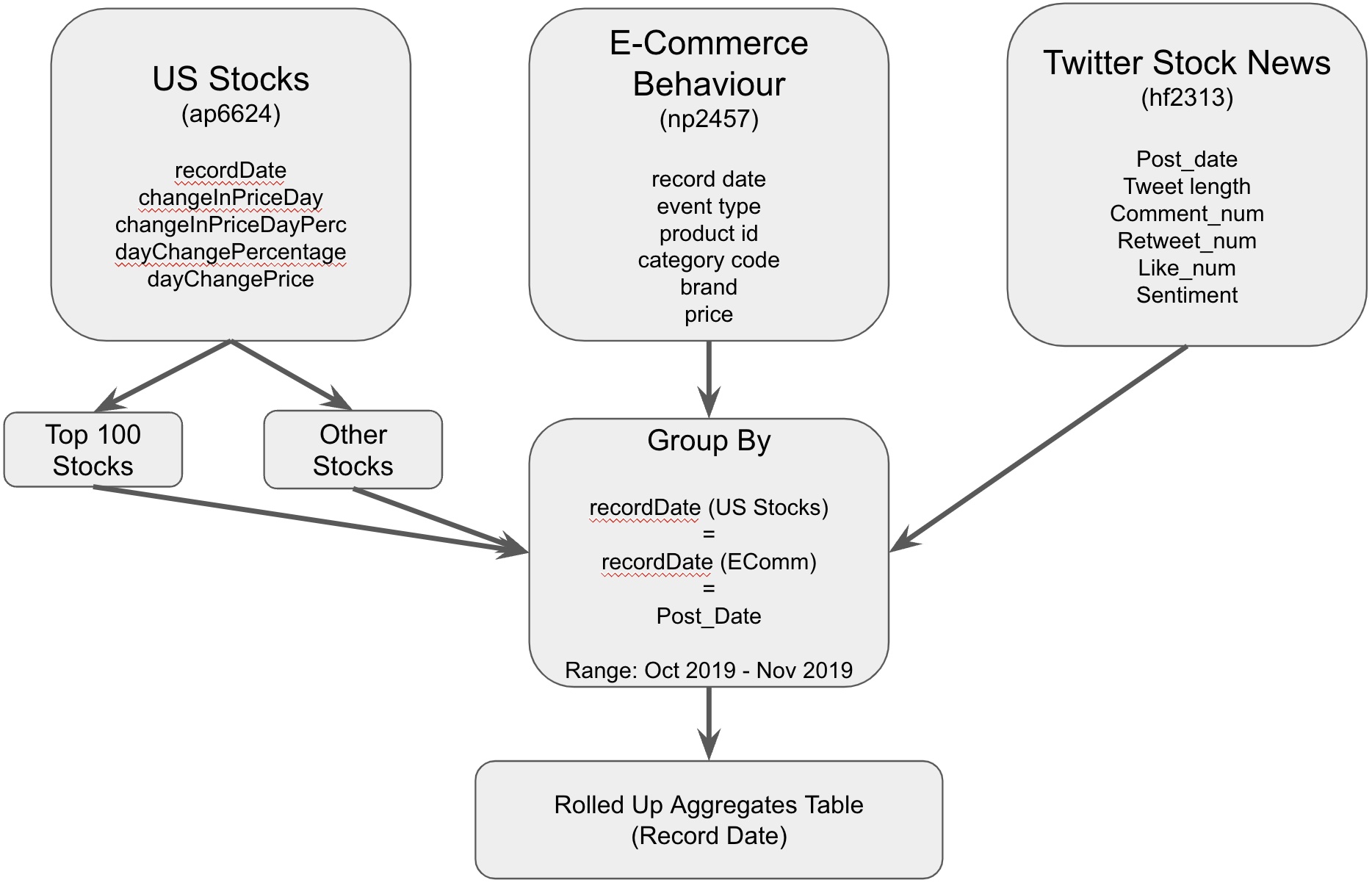}
  \caption{Datasets}
  \label{fig:datasets}
\end{figure*}

These are the datasets we used (shown in Figure~\ref{fig:datasets})
\subsection{US Stock Dataset}
The US Stock dataset\cite{us_stock_dataset} is a comprehensive collection of data on stocks and exchange-traded funds (ETFs) traded on the US stock market. It is updated on a daily basis and includes a wide range of information about these securities, including their prices, trading volumes, market capitalizations, and financial and economic indicators.

The dataset includes details such as the stock name, record date, open price, close price, stock split and we calculated the day-change price and day-change percentage. The stock name (which was obtained from the file name) and record date identify the specific stock being tracked, while the open and close prices provide information on the price at which the stock began and ended the trading day. The stock split indicates any changes to the number of outstanding shares of the security, which can affect the price and market capitalization. The day change price and day change percentage show the difference between the open and close prices and provide an indication of the security's performance over the course of the trading day.

In addition to these data points, the US Stock dataset also includes congressional transactions and multiple other financial statistics.

These data points can be useful for understanding the financial health and potential growth of the underlying companies and industries and for making informed investment decisions. Since the primary goal is to analyze trends, having a data-set with dates, well-formed values and precise data is important.

We split this dataset into the top 100 stocks vs the rest. \cite{top_100}
The combined size of this dataset was about 10GB.

\subsection{eCommerce Behavior Dataset}
The eCommerce dataset \cite{ecomm_dataset} used in this research consists of over 67 million values, with each row representing a specific event related to a product and a user. The events in the dataset are of various types, including view, cart, remove from cart, and purchase. The dataset includes several important columns, including an event time column that indicates when the event occurred, an event type column that specifies the type of event, a product ID column that identifies the product in question, a category code column that provides a meaningful name for the category (if available), a brand column that lists the brand name in lower case (if available), and a price column that lists the price of the product. 

These columns provide valuable information about the events in the dataset and are essential for understanding the relationships and trends present in the data. This dataset can be used to understand the preferences and behaviors of consumers across a wide range of categories and can be especially valuable for businesses looking to optimize their eCommerce strategies. \cite{Suguna2016BigDA}

During the research, we identified a few abnormalities in the data. Specifically, the data for November 18th was missing, and the data for November 19th was significantly larger than any of the other data points. We speculate that this may be due to the missing data from November 18th being included on November 19th. In order to account for these abnormalities, we have normalized the data for these two days in order to ensure that our analysis is accurate and reliable. 

\subsection{Stock Tweets Dataset}
The Stock Tweets Dataset \cite{stock_tweet_dataset} consists of data pertaining to stock-related tweets collected between 2015 and 2020. It was developed to identify potential stock market speculators and influencers. It can also be used to identify stock market trends. This dataset is in CSV format. 

This dataset is approximately 770 MB in size and contains around 3 million tweets. All tweets in the dataset are in English. Each record of the dataset contains the following information: tweet ID, handle of the twitter user who wrote the tweet, the timestamp of when the tweet was written, the actual tweets’ content, the number of comments the tweet received, the number of retweets of the tweet, and the number of likes on the tweet. The number of likes, comments and retweets can provide insights on twitter user engagement on stock related tweets. 

The Stock Tweets Dataset can be useful in identifying stock market patterns by analyzing the number of tweets, their sentiments, and user engagement. We can also use tweets about stocks while making financial decisions. This dataset will be used to examine the relationship between tweet sentiment and stock prices. \cite{mci/Nofer2015}

\section{Design and Architecture}
\begin{figure}[h]
  \includegraphics[width=\linewidth]{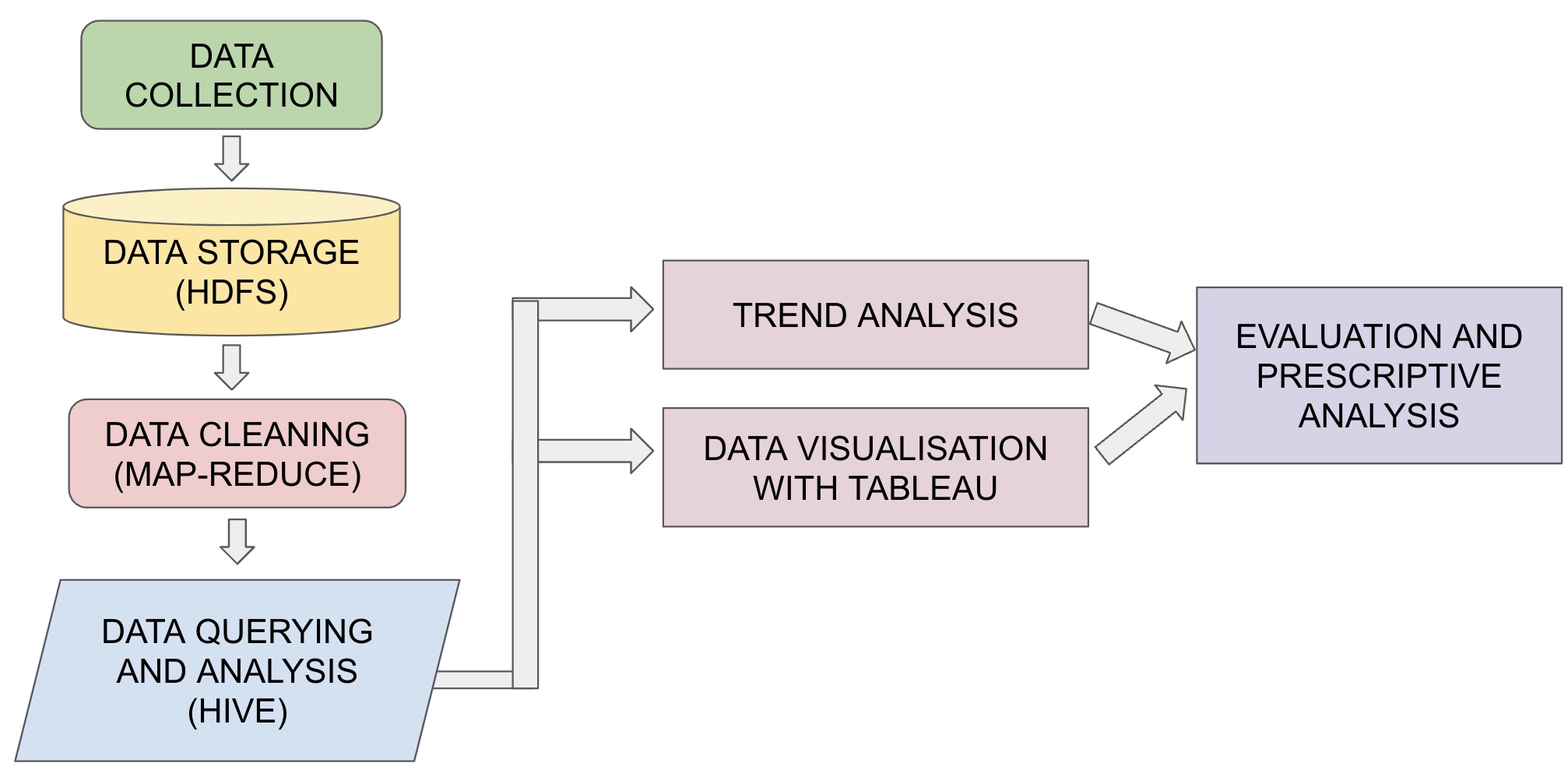}
  \caption{Data Pipeline}
  \label{fig:pipeline}
\end{figure}

We follow the data pipeline shown in Figure~\ref{fig:pipeline}. Each step is detailed in the following subsections.

\subsection{Data Storage using HDFS}
To begin with, we gathered consumer reviews from three datasets using the Kaggle API as mentioned in \hyperref[sec:dataset]{Section 4}. Although the datasets we acquired were in decent shape, we had to perform additional cleaning and filtering to make them suitable for our task and analysis. We stored all the data, totaling around 16 GB, on HDFS for fast access. To clean the data, we used the Map-Reduce in Java which is designed for distributed processing of large data sets across multiple computers. Map-Reduce is efficient and effective for handling large amounts of data, and it allowed us to quickly and easily clean and filter the data in our 3 datasets.

\subsection{Data cleaning using MapReduce}
As there were 3 different datasets, each with its features and data points, we did not want to merge these datasets (as shown in Figure~\ref{fig:datasets}). After evaluating each of the datasets, we decided to include the following columns for each - 
\begin{itemize}
    \item US Stock Data
    \begin{itemize}
        \item stockName
        \item recordDate
        \item openPrice
        \item closePrice
        \item stockSplit
        \item dayChangePrice
        \item dayChangePercentage
    \end{itemize}
    \item Stock Tweets Data
    \begin{itemize}
        \item Post\_date
        \item Tweet length
        \item Comment\_num
        \item Retweet\_num
        \item Like\_num
        \item Sentiment
    \end{itemize}
    \item E-Commerce Behavior Data
    \begin{itemize}
        \item record date
        \item event type 
        \item product id 
        \item category code
        \item brand
        \item price
    \end{itemize}
\end{itemize}

The next step was to clean the data. All our datasets are individual files were processed and stored in a CSV format. We used MapReduce to clean the data and performed the following steps:
\begin{itemize}
    \item For all data, we trimmed any leading spaces and quotes.
    \item Filtered out all the extra columns and rows containing null values and noisy data
    \item For the tweet text, we performed the following steps:
    \begin{itemize}
        \item Removed non-alphabetic characters
        \item Removed digits
        \item Removed stopwords. Words like a, an, the, and so on are considered to be stopwords. This is a common step in Language Analysis and it usually improves performance.
        \item Extracted the sentiment from the tweet.
    \end{itemize}
    \item To get the sentiment for the tweets, we used TextBlob \cite{loria2018textblob}. TextBlob is a Python library that allows you to perform various natural language processing (NLP) tasks, including part-of-speech tagging, noun phrase extraction, sentiment analysis, classification, translation, and others. It is compatible with both Python 2 and 3 and offers a simple API for accessing these functions.
    \item Review date was processed to get to the uniform form yyyy-mm-dd from varying formats.
    
\end{itemize}

In this process, the Mapper takes a single row of the CSV as input and performs the series of above mentioned steps on the data. Once the Mapper has cleaned the data, it sends a string of the relevant columns, separated by commas, to the Reducer.
The Reducer receives the cleaned data from the Mapper and writes it to the output file on HDFS i.e. writes the key to HDFS. This output file can then be easily accessed and queried by Hive

It's important to note that only the sentiment analysis was done using Python in this process.

\begin{figure}[h]
  \includegraphics[width=\linewidth]{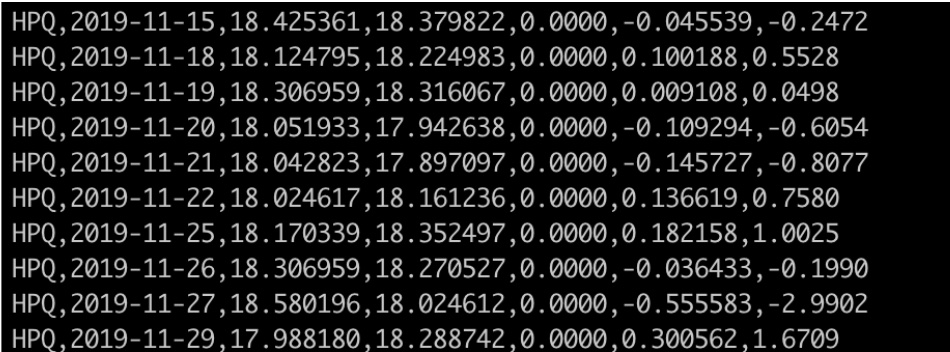}
  \caption{Screenshot of the Cleaned Stock Dataset}
  \label{fig:stock_dataset}
\end{figure}
\begin{figure}[h]
  \includegraphics[width=\linewidth]{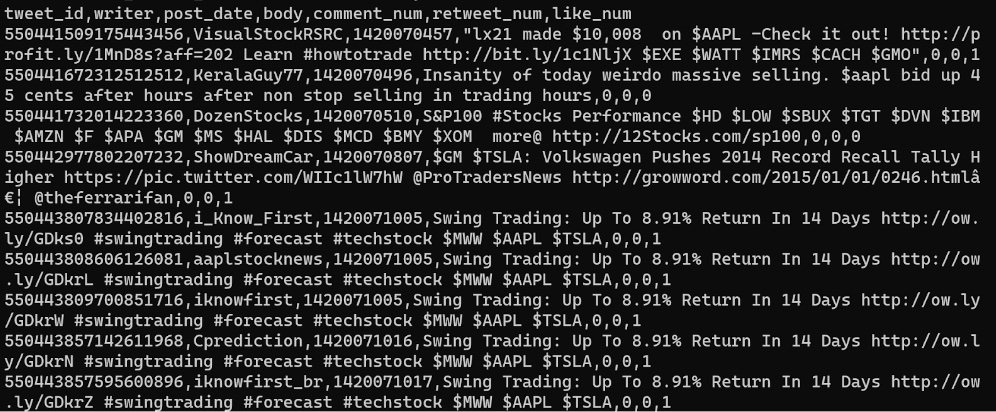}
  \caption{Screenshot of the Cleaned Tweet Stock Dataset}
  \label{fig:tweet_dataset}
\end{figure}
\begin{figure}[h]
  \includegraphics[width=\linewidth]{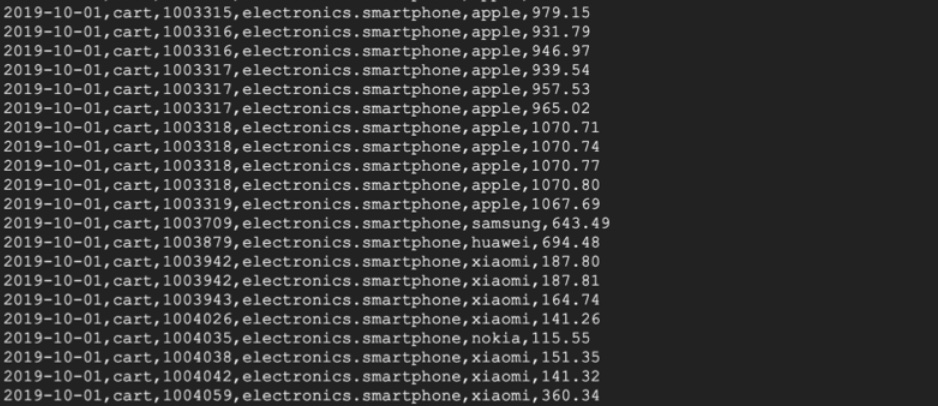}
  \caption{Screenshot of the Cleaned E-Commerce Behavior Dataset}
  \label{fig:ecomm_dataset}
\end{figure}

\subsection{Data Querying and Analysis using Hive}
Using Map-reduce to clean the data results in multiple output files, which we then fed into Hive for further analysis. Hive is a data warehousing tool that allows users to work with large datasets stored in distributed storage using SQL. It is particularly well-suited to our needs because it supports a relational schema, which is appropriate for our dataset. We also considered using technologies like MySQL and HBase, but ultimately chose not to use them due to various limitations. Specifically, MySQL does not scale well and cannot directly load data from Map-reduce, while our dataset is more suited to a relational form than a wide columnar storage, which ruled out the use of HBase.

For all the datasets, three initial external tables were created. For further analysis, we created multiple new tables and views which helped in reducing the time of running queries with multiple joins, group-bys and filter.

Figures~\ref{fig:tweet_dataset},~\ref{fig:stock_dataset},~\ref{fig:ecomm_dataset} are screenshots of the cleaned data from the dataset. 

\subsection{Data Visualization using Tableau}
Tableau is a popular business intelligence and data visualization tool that allows users to easily create interactive charts, graphs, and dashboards from various data sources. Hive can conveniently be used as one of the data sources. Tableau can connect to Hive using a native connector or through a connection to a SQL-based database that acts as a gateway to Hive. We used the Cloudera Hadoop driver for connecting them.

Once connected, we used Tableau's drag-and-drop interface to build visualizations of their data and perform various analyses. We also used the Tableau options for filtering, grouping, and aggregating data, as well as the ability to create calculated fields and custom SQL queries
\\
Most of our analysis involved joining our 3 tables on the \texttt{Date} field. For that, we made use of Relationships instead of Joins in Tableau.

A relationship is a way to link two data sources together without actually combining the data. Instead of creating a new dataset, a relationship allows you to use fields from both data sources in your visualization and analysis, as if they were a single dataset and can be used to filter and aggregate data across both data sources. Relationships were useful as we wanted to preserve the independence of the data sources and didn't want to combine the large datasets.

\section{Results}

\subsection{Category Distribution and Purchase Trends}
\begin{figure}[h]
  \includegraphics[width=\linewidth]{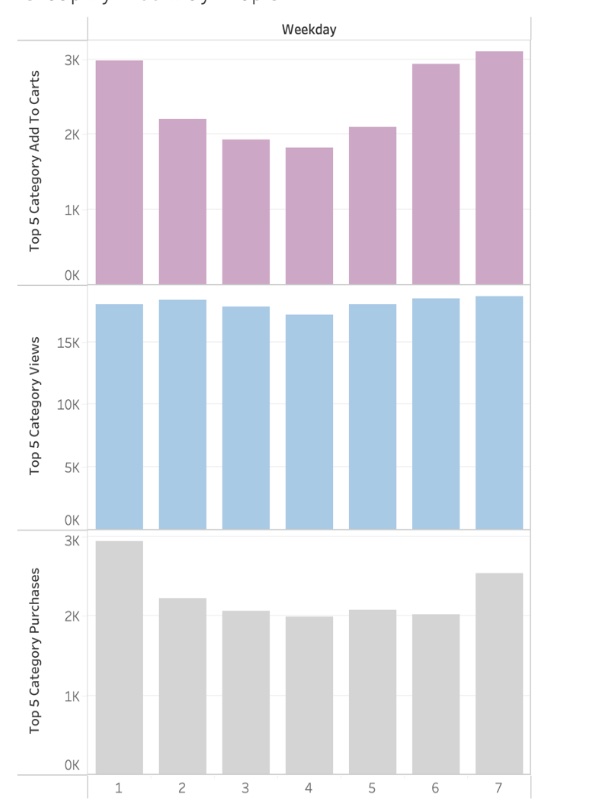}
  \caption{Top Categories}
  \label{fig:topactions}
\end{figure}

An analysis of the purchase categories in our e-commerce data revealed that the most popular categories are smartphones, shoes, appliances, and TVs. We also observed that the "add to cart" and "purchase" events are significantly higher on the weekends, specifically on Saturday and Sunday. This aligns with the idea that users have more free time to shop online on these days (see Figure~\ref{fig:topactions}).

\begin{figure}[h]
  \includegraphics[width=\linewidth]{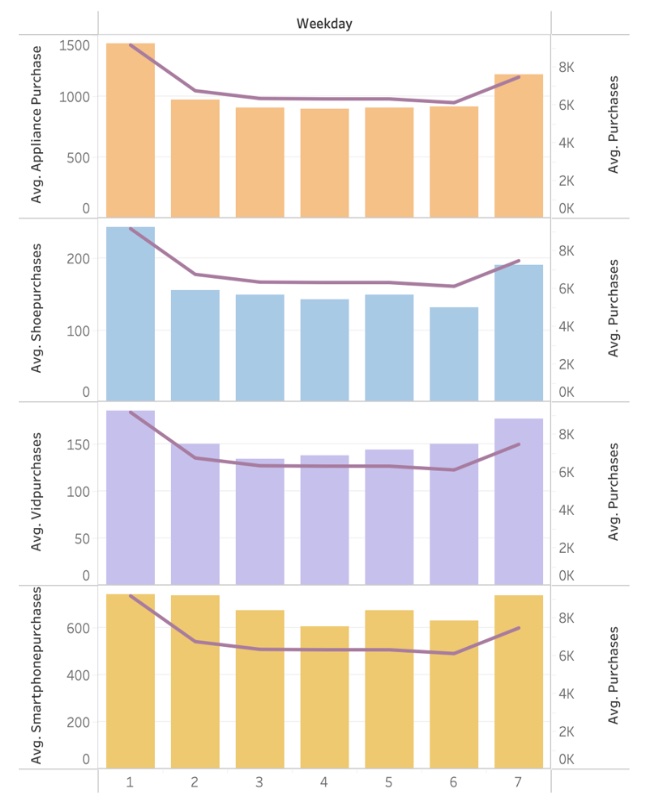}
  \caption{Top Purchase Categories}
  \label{fig:TopPurchaseCategories}
\end{figure}

The plot, shown in Figure~\ref{fig:TopPurchaseCategories}, demonstrates that the add to cart trend holds true for categories such as TVs, shoes, and appliances, which are related to apparel and home products. However, this trend does not hold for smartphones, which are the top purchase category. This could be because smartphones are more likely to be purchased on a need basis or during times of peak sales, rather than just based on leisurely browsing.

\subsection{Weekly Twitter User Engagement}
\begin{figure}[h]
  \includegraphics[width=\linewidth]{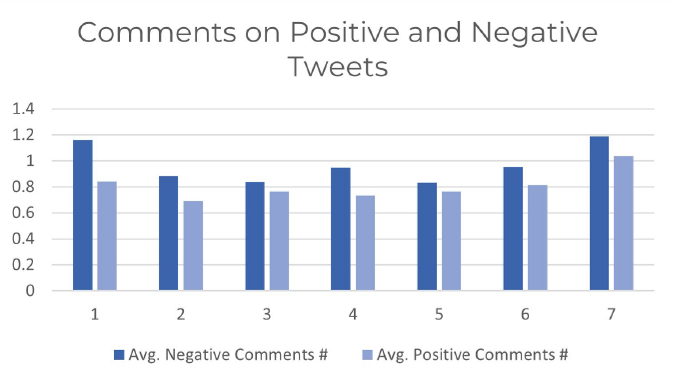}
  \caption{Comments on Positive and Negative Tweets}
  \label{fig:WeeklyComments}
\end{figure}

\begin{figure}[h]
  \includegraphics[width=\linewidth]{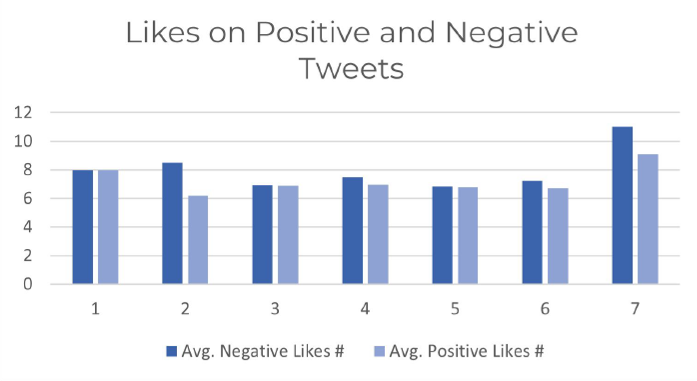}
  \caption{Likes on Positive and Negative Tweets}
  \label{fig:WeeklyLikes}
\end{figure}

On looking at user engagement on tweets with positive and negative sentiments on different days of the week (see figures ~\ref{fig:WeeklyComments} and ~\ref{fig:WeeklyLikes}), we can see that the number of likes and comments on negative tweets is higher than those of positive tweets, showing that users are more likely to engage on tweets having negative sentiment. This could be due to the fact that users tend to be more critical on social media. This could also be corroborated by the fact that usually negative news spreads faster on Twitter \cite{negative_news_spread_faster}. 

\subsection{Purchases and Positive Stocks from the Previous Day}
\begin{figure}[h]
  \includegraphics[width=\linewidth]{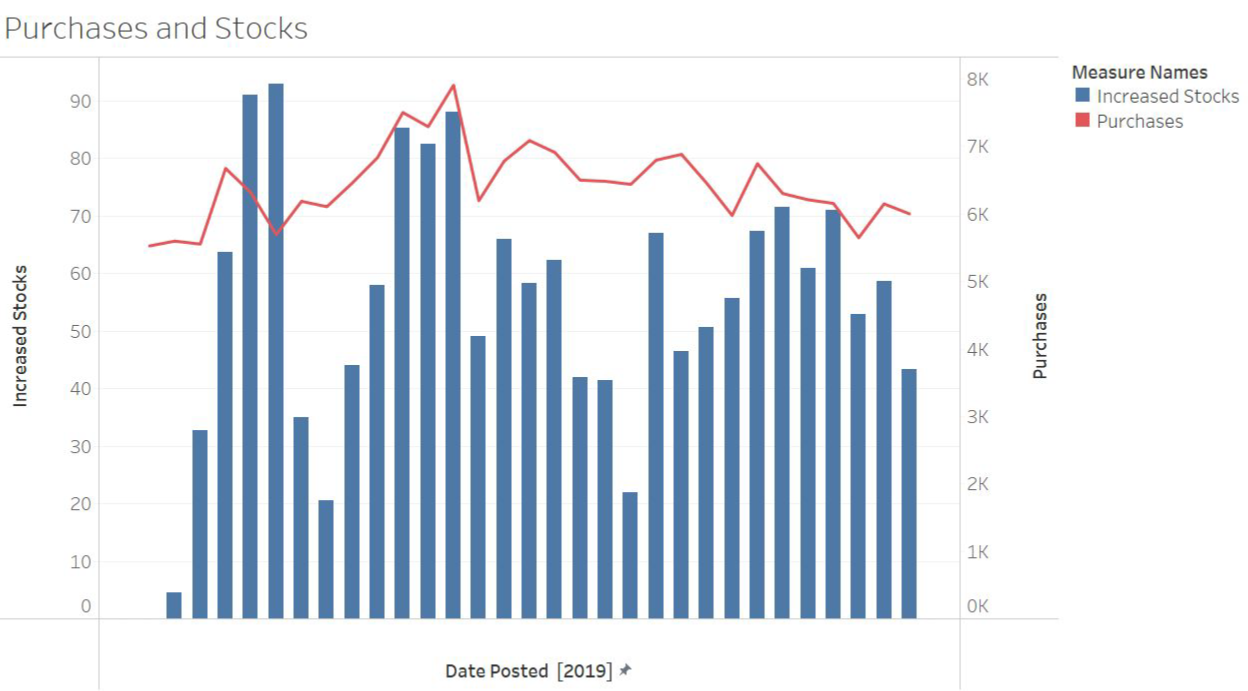}
  \caption{Comparing Purchases and Stocks}
  \label{fig:PurchasesAndStocks}
\end{figure}
We plot the number of purchases made on a given day against the number of stocks that increased in value on the previous day (see figure ~\ref{fig:PurchasesAndStocks}). From the plot, it is clear that there is a trend such that when the number of purchases increases, the number of stocks that increased in value also increases. Similarly, when there is a decrease in the number of stocks that increased in value, we see a corresponding drop in the number of purchases, although the drop is not as severe.

It appears that when more purchases are made, there is a higher likelihood that more stocks will increase in value, and vice versa. This could be due to various factors, such as changes in investor sentiment or market conditions, which may influence both the volume of purchases and the performance of stocks.

\subsection{Positive Tweet Trending with Well Performing Stocks}
In this study, we introduce the terms "big winners" and "big losers" to refer to stocks that have experienced significant changes in price. Specifically, we define "big winners" as stocks that have increased in value by more than 5\% and "big losers" as stocks that have decreased in value by more than 5\%. These terms are used to group stocks based on the magnitude of their price changes and to analyze the factors that may contribute to these changes. 

By classifying stocks in this way, we aim to gain insights into the performance of different types of stocks and identify trends or patterns that may be relevant to investors.

\subsubsection{Positive Tweet Trending with Well Performing Stocks}
\begin{figure}[h]
  \includegraphics[width=\linewidth]{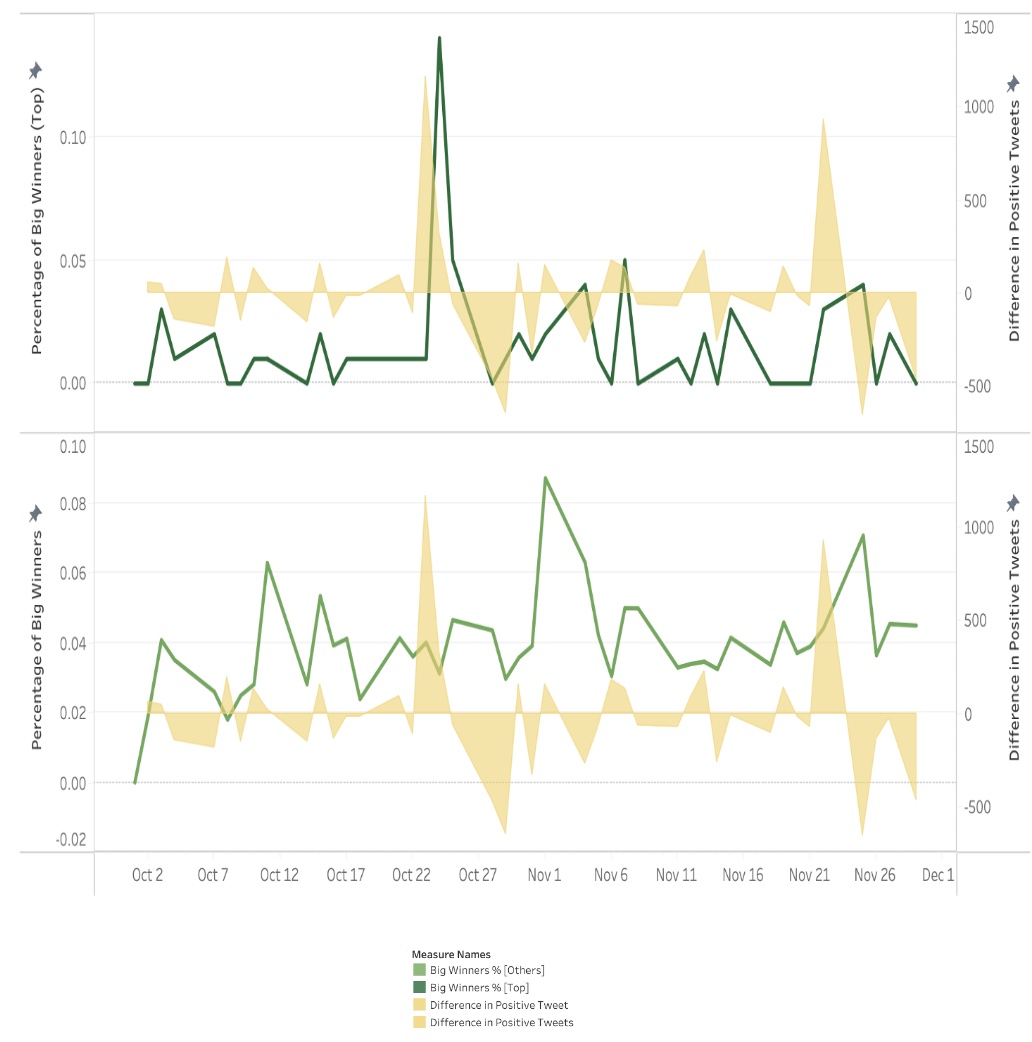}
  \caption{Positive Tweets Trending with Well Performing Stocks}
  \label{fig:pos_good_stock_weekly}
\end{figure}

When comparing the number of positive tweets on a given day to the number of positive tweets on the previous day (see figure ~\ref{fig:pos_good_stock_weekly}), we see that there is a high level of similarity between the number of "Big Winning" stocks on the day. We also observe a general lag of 0-1 days, which suggests that an increase in the number of tweets is a reliable indicator of an increase in the number of "Big Winning" stocks in the coming day or two. However, we do not see a clear similar trend for the rest of the stocks.

These findings suggest that the general sentiment expressed in tweets is heavily biased towards the top stocks. It appears that tweets about these stocks are more likely to predict future performance, while tweets about the other stocks are less indicative of future performance. This suggests that these tweets may be a useful tool for predicting the performance of top stocks, but may not be as reliable for predicting the performance of the rest of the stocks.

\subsubsection{Analysing Positive Tweets and Good Stocks on a Week Number basis}
In addition to the analysis described above, another visualization that supports our findings is the correlation between the change in the number of positive tweets (grouped by week number) and the percentage of "Big Winners".  

\begin{figure}[h]
  \includegraphics[width=\linewidth]{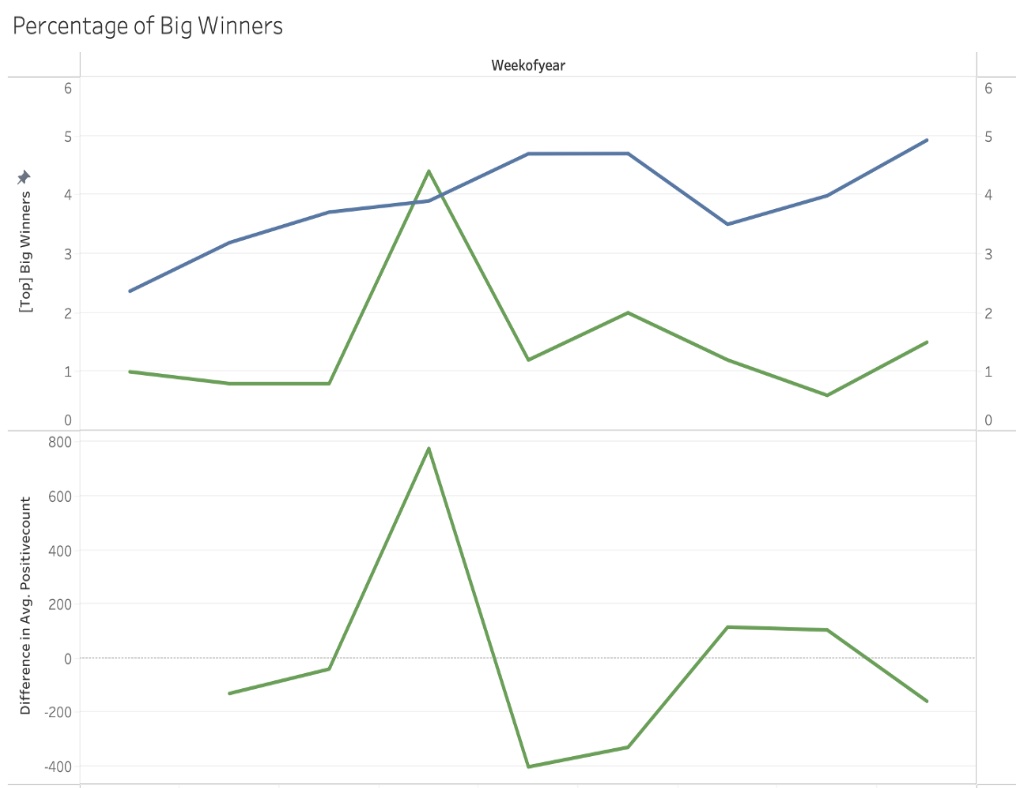}
  \caption{Positive Tweets and Good Stocks on a Week Number Basis}
  \label{fig:pos_good_stock_weekno}
\end{figure}

We noticed a spike in both the tweet count and the number of big winners in the second-to-last week of October (see figure ~\ref{fig:pos_good_stock_weekno}). Upon further investigation, we discovered that this was when most companies announced their end-of-quarter results \cite{q_res}, which were generally positive. This reinforces the idea that tweets can serve as useful indicators of market trends and may be valuable for predicting the performance of stocks.

\subsection{Comparing Top Stocks vs All Other Stocks}
\begin{figure}[h]
  \includegraphics[width=\linewidth]{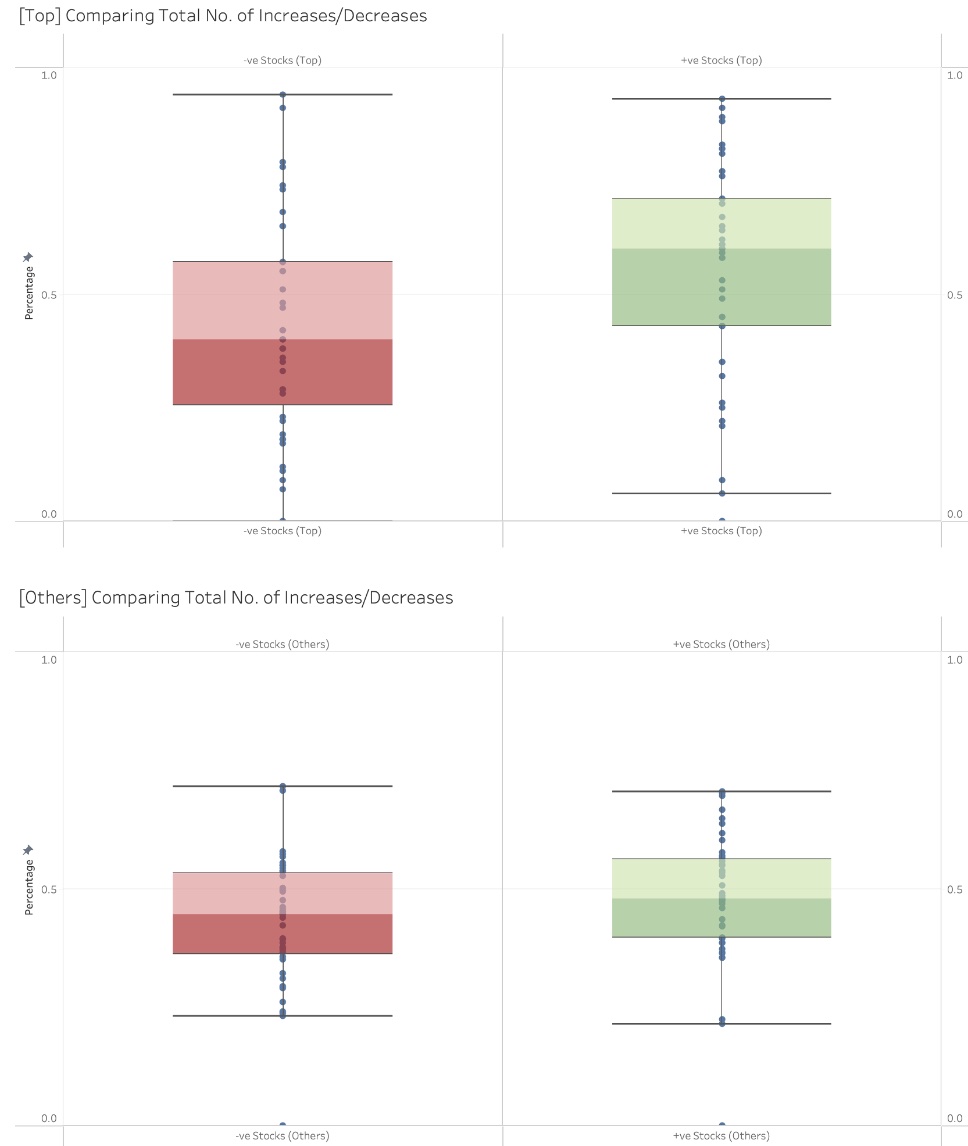}
  \caption{General Trends of Top Stocks vs Rest}
  \label{fig:top_v_rest}
\end{figure}

When examining the percentage of stocks that are increasing (shown in red) and decreasing (shown in green), we see that the spread is much wider for the "top" stocks (see fig ~\ref{fig:top_v_rest}). This indicates that when the market trends in a positive direction, a large majority of top stocks tend to move in the same direction. The median positive movement is around 60\%. On the other hand, the percentage of stocks with a negative movement is more limited, at around 40\%. This suggests that even on bad days, the negative performance of top stocks is somewhat contained.

In contrast, when we look at the other stocks, we see that the spread is narrower. This means that there are fewer days where all of the other stocks move in either a positive or negative direction. Overall, these observations suggest that the top stocks are more strongly influenced by market trends, while the other stocks are more resistant to these trends.

\subsection{Stock Market based on the Day of the Week}
When we roll up our stock data by weekday (Monday-2 to Friday-5), we see that there are some similarities between the behavior of the "top" stocks and the rest of the stocks. For example, we observe that there is a noticeable difference in the number of "big winners" and "big losers" on Fridays. This pattern is also evident when we look at the data without grouping it, which suggests that it is a consistent trend.

\begin{figure}[h]
  \includegraphics[width=\linewidth]{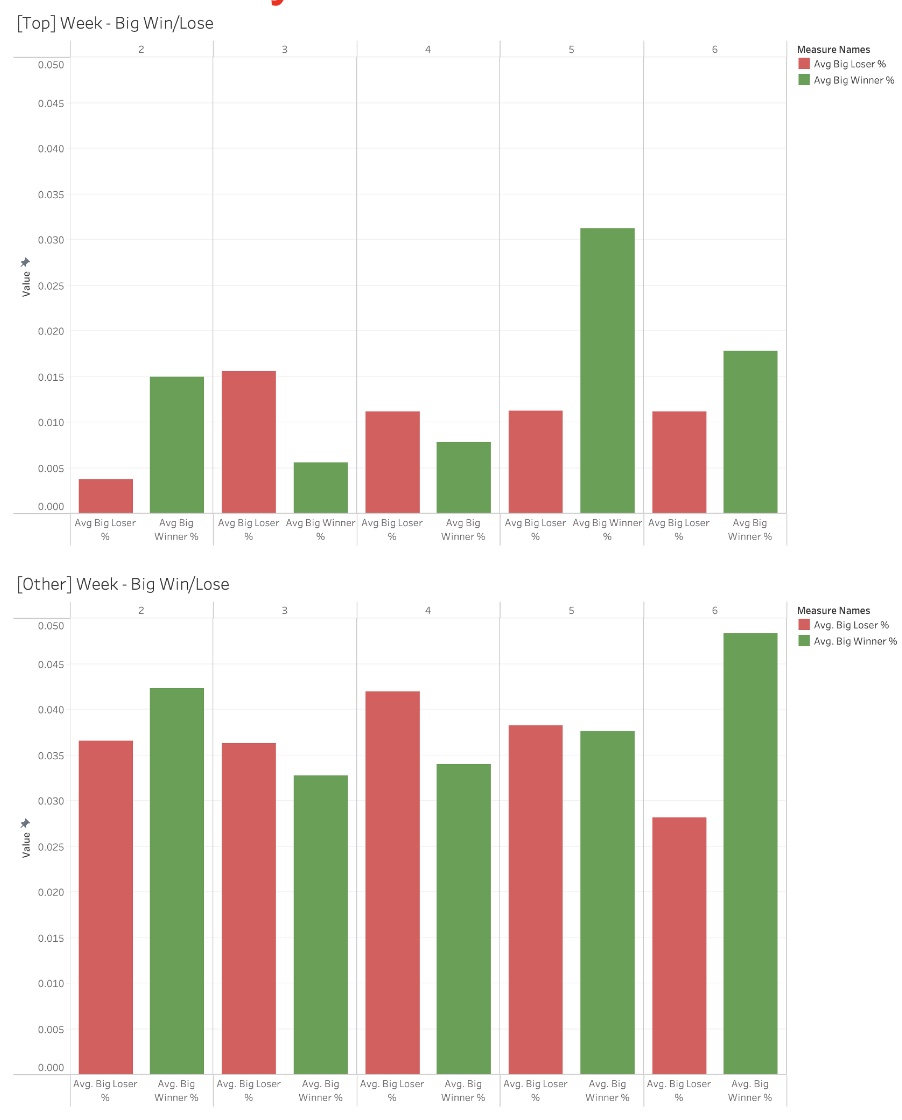}
  \caption{Comparing Top Stocks with the rest based on the Day of the Week}
  \label{fig:big_win_lose}
\end{figure}

Furthermore, we notice that the top stocks tend to be less volatile, meaning that they experience fewer large changes in price. When we compare the relative number of stocks that increase or decrease by more than 5\% on a given day, we see that this number is lower for the top stocks than it is for the other stocks. This could be due to the fact that top stocks are generally larger and more established companies, which tend to be more stable and less prone to dramatic price fluctuations (see figure ~\ref{fig:big_win_lose}). 

These observations suggest that the behavior of the top stocks is somewhat different from the rest of the stock market, which are worth considering when making investment decisions.

\section{Conclusion}
In conclusion, we utilized various big data tools, such as Hadoop, Map-Reduce, Hive, and Tableau, to perform sentiment, correlation, and trend analysis on e-commerce, stock, and tweet data. We demonstrated the potential for analyzing stocks, tweets, and eCommerce data in order to understand correlations and trends in the market. By performing various forms of analysis, including sentiment analysis and categorical analysis, we were able to gain valuable insights into the data. 

The use of Hadoop for data storage and Hive for querying, as well as the creation of custom columns and features, allowed for efficient and effective analysis. The results of this analysis were effectively visualized using Tableau, providing a clear and comprehensive overview of the data. Overall, this highlights the importance of data analysis in the modern business world and the potential for using multiple datasets to gain a more comprehensive understanding of market trends.

\section{Future Scope}
We believe there are several potential avenues for expanding upon this research in the future. One possibility is to include a wider time series or range of data, which could provide a more comprehensive understanding of the trends and correlations present in the data. This could also allow for a more in-depth analysis of the overlap between the three datasets. Another possibility is to include tweets about current events and news, in addition to those related to stocks, which could provide additional context and potentially reveal new insights. Additionally, the eCommerce dataset could be expanded to include a greater number of and more distinct categories, which could provide a more complete picture of the relationships between eCommerce data and other variables. Finally, gathering stock data related to non-US stocks could provide a more global perspective and allow for a comparison of trends and correlations across different markets. 

All of these potential avenues for future research have the potential to deepen our understanding of the relationships and trends present in the data and provide valuable insights.
\Urlmuskip=0mu plus 1mu\relax
\bibliographystyle{unsrt}
\bibliography{bibflile}

\end{document}